\documentclass[
twocolumn,showpacs,preprintnumbers,amsmath,amssymb,
superscriptaddress,unsortedaddress]{revtex4}

\usepackage{graphicx}
\usepackage{dcolumn}
\usepackage{bm}
\newcommand{\deff}{d_{\textrm{eff}}}
\newcommand{\Deff}{D_{\textrm{eff}}}

\begin{document}


\title{Quasiperiodic dynamics of coherent diffusion: a quantum walk approach}

\author{Antoni W\'ojcik}
\affiliation{Faculty of Physics,
 Adam Mickiewicz University,
Umultowska 85, 61-614 Pozna\'{n}, Poland.
}
\author{Tomasz {\L}uczak}
\altaffiliation{Corresponding author. 
Phone: +48 (61) 829-5394, 
fax: +48 (61) 829-5315. 
E-mail: tomasz@amu.edu.pl}
\affiliation{Faculty of Mathematics and Computer Science,
Adam Mickiewicz University,
Umultowska 87, 61-614 Pozna\'{n}, Poland.
}
\author{Pawe{\l} Kurzy\'nski}
\affiliation{Faculty of Physics,
 Adam Mickiewicz University,
Umultowska 85, 61-614 Pozna\'{n}, Poland.
}
\author{Andrzej Grudka}%
\affiliation{Faculty of Physics,
 Adam Mickiewicz University,
Umultowska 85, 61-614 Pozna\'{n}, Poland.
}
\author{Ma{\l}gorzata Bednarska}
\affiliation{Faculty of Mathematics and Computer Science,
 Adam Mickiewicz University,
Umultowska 87, 61-614 Pozna\'{n}, Poland.
}

\date{July 15, 2004}%

\begin{abstract}
We study the dynamics of a generalization of quantum coin walk on the line
which is a natural model for a diffusion modified by quantum
or interference effects. 
In particular, our results provide surprisingly simple explanations to
phenomena observed by Bouwmeester {\em et al.}  
(Phys.\ Rev.\ A {\bf 61}, 13410 (1999))
in their optical Galton board experiment,
and a  description of a stroboscopic quantum walks
given  by Buershaper and Burnett (quant-ph/0406039) 
through numerical simulations.
We also provide heuristic explanations for the behavior of our  
model which show, in particular, that its dynamics can be viewed as 
a discrete version of Bloch oscillations.
\end{abstract}

\maketitle

The behavior of quantum analogs of classical systems with 
chaotic dynamics has recently attracted much attention of both 
theoretical and experimental physicists. 
One of the most widely studied non-classical phenomena 
emerging in such systems are quantum suppression (dynamical localization) 
and enhancement (quantum resonance) predicted for various models of spectral diffusion. 
Both suppression and enhancement have been observed for a model 
of $\delta$-kicked rotor in an experiment in which cold atoms interact 
with a pulsed optical lattice [1--4] and,  recently, 
in superconducting nanocircuits~\cite{Mon}. 
Another approach to quantum chaos have been presented 
by Bouwmeester {\em et al.}~\cite{Bou}. 
They realized an optical version of a Galton board (OGB), 
and studied how the diffusion in OGB is influenced 
by a choice of two parameters -- the diabatic transition probability $D$ 
and the relative phase $\Phi$ (precise definitions of $D$ and $\Phi$ are given below).
The dynamics of  OGB turned out to be very different from a classical Galton board 
which corresponds to a standard random walk and results in Gaussian diffusion. 
In particular, the authors of the OGB experiment concluded that  
``suppression of diffusion can occur in the form of almost perfect 
recurrences of the initial level population''~\cite{Bou}. 
Such an effect had been already predicted by Harmin \cite{Har} 
in the adiabatic limit ($D\ll 1$), but, since these recurrences 
must vanish in the strictly diabatic limit ($D=1$), 
it came as a surprise that they were observed even for the transition 
probability $D$ as high as $0.8$. 
Prompted by these observations the authors speculated that the system 
has two different modes of behavior, adiabatic and diabatic. 
A similar picture of the dynamic of OGB was presented by T\"orm\"a \cite{Tor} 
who even provided some heuristic arguments suggesting that, 
in fact, the transition between adiabatic and diabatic phases 
may be similar to the phase transition in two-dimensional Ising model. 
We also remark that in the OGB experiment the diffusion is suppressed 
for rational values of the relative phase $\phi/2\pi$, 
for which one would rather expect an enhancement caused by quantum resonance
effects. 

In this paper we model coherent diffusion using a generalized coin 
quantum walks (GCQW) providing a consistent, although somewhat unexpected, 
explanation for recurrences phenomena observed in the OGB experiment. 
In particular, our model predicts that the recurrences changes with 
the transition probability $D$ in a smooth, continuous way which 
refutes the conjecture that in the dynamics of such a system 
one can distinguish two different adiabatic and diabatic modes of behavior.  
Moreover, our approach explains nicely why and when we can observe recurrences 
and ballistic diffusion; it also predicts some new multiple recurrences effects 
which, most likely, can be verified experimentally within current technology. 
Besides of the quantitative analysis we offer a simple intuitive explanation 
of basic features of coherent diffusion dynamics which explores similarities 
between GCQW and Bloch dynamics. 
Let us also remark that the quantum walk framework and results presented here 
are not restricted to particular implementation of coherent diffusion such as OGB. 
In fact, we believe that our approach can capture many general properties of 
diffusion modified by quantum or interference effects. 

The optical Galton board (OGB) studied by Bouwmeester {\em et al.}~\cite{Bou} 
consisted of a ladder of equally spaced levels periodically coupled to each other 
via Landau-Zener crossings. Due to the structure of equally spaced levels OGB 
resembles a $\delta$-kicked harmonic oscillator. 
However, as was pointed by Bouwmeester {\em et al.} \cite{Bou}, it can be 
assumed that Landau-Zener crossings induce transitions between 
neighboring levels only. 
Thus, the OGB experiment can be as well considered as an implementation 
of a coined quantum walk \cite{Kni}. 
Let us recall that a coined quantum walk (CQW) is a quantum analog of 
a random walk on graph proposed by Aharonov, Davidovich and Zagury \cite{ADZ}, 
extensively studied for the last few years in the hope of constructing 
efficient quantum algorithms~(see \cite{Kem}).  
Buerschaper and Burnett~\cite{Bue} have noticed that CQW can also be useful for 
modeling and studying quantum chaos. 
Here we follow this approach and characterize the coherent diffusion in the terms 
of generalized CQW. 
In a model of CQW on the line~\cite{Aha} the evolution of the system is determined 
by a repeating action of a unitary operator $U$ defined on a tensor product 
of two Hilbert spaces $H_c \otimes H_n$. The base of $H_c$ consists of the coin 
states $|c\rangle$, ($c=0,1$), whereas $H_n$ is spanned by vectors  
$|n\rangle$, ($n\in \mathbb Z$), which correspond to the position of 
a particle on the line. In a single step of the evolution of the system 
we toss a coin, i.e., change the coin state $|c\rangle$, and move the particle 
to one of the two neighboring state of the line determined by $|c\rangle$. 
Let 
$U= \big(\sum_{c}|c\rangle\langle c|\otimes S_c\big)(C\otimes I)$, 
where $C\in SU(2)$ is an arbitrary coin tossing operator, and $S_c$ 
is a translation operator defined as $S_c|n\rangle=|n+(-1)^c\rangle$. 
The state vector $|\Psi\rangle$ of the system evolves in (discrete) time $t$ 
according to  $|\Psi(t)\rangle=U^t|\Psi(0)\rangle$. 
Note that if the state of the coin is measured after each step, 
our model corresponds precisely to the classical random walk. 

If a quantum walk is to describe OGB it must take into account the phase relations, 
so the above mentioned basic model has to be generalized. 
This can be done by replacing $U$ by 
$U_\phi=\big(\sum_{c}|c\rangle\langle c|\otimes S_{c\phi}\big)
(C\otimes I)$, where $S_{c\phi}|n\rangle=e^{i\phi(n)}|n+(-1)^c\rangle$, 
and $\phi(n)$ is a phase acquired by the particle at the position $n$. 
We remark that the generalized coin quantum walk (GCQW) 
defined above is consistent with a recently proposed 
stroboscopic quantum walk (SQW)~\cite{Bue}. 
In the SQW model a particle walks on a space consisting of 
eigenstates of some Hamiltonian $H$, 
and the evolution is periodically perturbed with an operator $U$.  
Both SQW and CQW models can be obtained from GCQW by taking  
$\phi(n)=\langle n|H|n\rangle t_p$ (where $t_p$ is a period 
of perturbation), and $\phi(n)=const$,  respectively. 
Below we concentrate on harmonic case $\phi(n)=n\Phi$, 
where, furthermore, $\Phi$  is rational fraction of $2\pi$, 
i.e., $\Phi=2\pi\tfrac{q}{p}$, with coprime $q$ and $p$ 
(although we shall make some assertions about the irrational case as well).
In order to study GCQW we  assume that a particle walks not on the line 
but on a long cycle of length $N$, where $N$ is a large multiplicity of $p$ 
so that $\phi(N)=\phi(0)$. 
Thus, here and below $n\in \mathbb Z_N$  
and addition in the definition of $S_{c\phi}$ is taken modulo $N$.  
Obviously, for times $t<N/2$, the behavior of both the line and the $N$-cycle 
models is identical.
 
 	Let a coin tossing operator be in the form 
$C=\sum_c\big((-1)^c d|c\rangle\langle c|+a |c\rangle\langle 1-c|\big)$
(if $D$ is the diabatic transition probability in the OGB experiment, then 
$d=\sqrt{D}$ and $a^2+d^2=1$). 
For the eigenvalues $r_{jus}$,  ($j=0,1,\dots,N/p-1$; $u=0,1$; $s=0,1,\dots,p-1$), 
of the unitary operator $U_\phi$, for the $N$-cycle model we have got 
$r_{jus}=\omega_p^s z_{ju}$  for an odd $p$, 
and $r_{jus}=\omega_p^s (2i\lambda_j z_{ju}-1)$  for an even $p$,  
where $\omega_p=e^{2\pi i/p}$,
$z_{ju}=\big((-1)^u\sqrt{1-\lambda_j^2}-i\lambda_j\big)^{1/p}$,
$\lambda_j=\deff\sin\tfrac{\pi T j}{N}$, 
the effective amplitude is given by $\deff=d^{T/2}$, 
and $T=p$ for an even $p$ while $T= 2p$ if $p$ is odd.

The probability that the particle returns to its initial state is given 
by $P(t)=|\langle\Psi(t)|\Psi(0)\rangle|^2$. 
Let $f(t)\equiv \langle\Psi(t)|\Psi(0)\rangle=\sum_{j,u,s}A_{jus}r_{jus}^t$, 
where $A_{jus}=|\langle jus|\Psi(0)\rangle|^2$, and $|jus\rangle$ 
is the eigenvector corresponding to the eigenvalue $r_{jus}$. 

In the adiabatic limit, when $d\to0$, we have $r^T_{jus}=\pm 1$ 
independently of $j$, $u$, and $s$, 
so in this case $P(T)=1$ as predicted by Harmin~\cite{Har}.  For $d>0$ we have
\begin{equation}\label{eq1}
f(T)=\sum_j A^+_j \textrm{Re}(\zeta_ju)+\sum_j A^-_j\textrm{Im}(\zeta_{ju})\,
\end{equation}
where 
$\zeta_{ju}\equiv r^T_{jus}=\pm \big(1-2\lambda_j^2-
2i(-1)^u\lambda_j\sqrt{1-\lambda^2_j}\big)$  
(the sign depends on parity of $p$), and $A^\pm_j=\sum_s(A_{j0s}\pm A_{j1s})$. 
Due to symmetries of eigenvectors second sum in (\ref{eq1}) vanishes, 
whereas $A^+_j=p/N$ for every $j=0,1,\dots,N/p-1$. Consequently, 
\begin{equation}\label{eq2}
P(T)=(1-\Deff)^2,
\end{equation}
where the effective probability of the diabatic transition is defined as
\begin{equation}\label{eq3}
\Deff\equiv \deff^2=D^{T/2}.
\end{equation}  

\begin{figure}
\scalebox{.8}
{\includegraphics{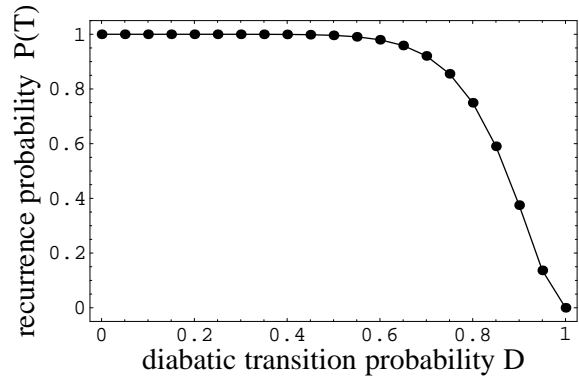}}
\caption{\label{f1} The recurrence probability $P(T)$ as a function of 
the diabatic transition probability $D$ for $p=9$. 
The curve is given by (\ref{eq2}), the dots are found by numerical simulations.}
\end{figure}

\noindent Note that, contrary to the suggestions that in the dynamic of OGB 
there exist two sharply distinguished adiabatic and diabatic phases \cite{Bou,Tor}, 
$P(T)$ is a smooth function of $D$ (see Fig.~\ref{f1}). 
In fact, for every given $D$, one can always take $p$ large enough 
so that $P(T)\cong 1$  and obtain an almost perfect recurrence, although, 
perhaps, due to decoherence effects,
for such a $p$ this phenomenon can hardly be verified experimentally.
 
\begin{figure}
\scalebox{.8}
{\includegraphics{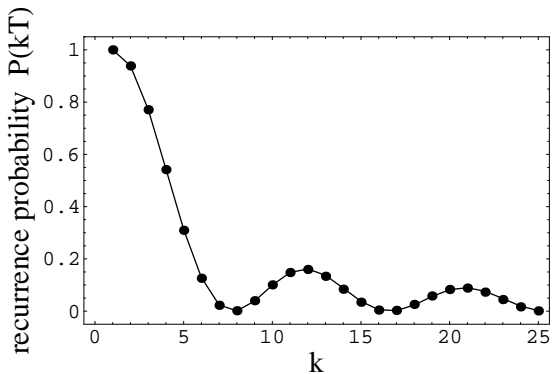}}
\caption{\label{f2} The probability of a multiple recurrence $P(kT)$ as a function of 
$k$ for $D=1/2$ and $p=10$. 
The curve is given by (\ref{eqkt}), 
the dots are found by numerical simulations.}
\end{figure}

\begin{figure}
\scalebox{.8}
{\includegraphics{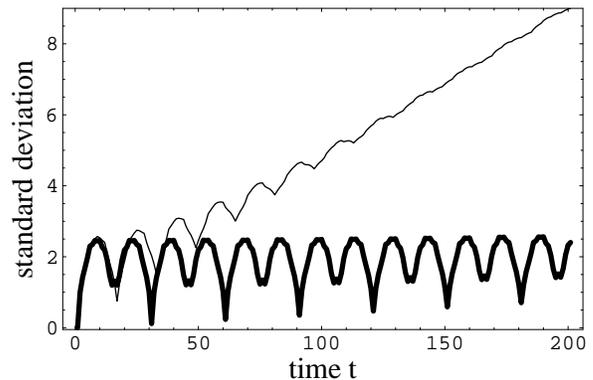}}
\caption{\label{f3}  The evolution of the standard deviation $\sigma$  for  
$q=1$, $p=16$ (thin line) and  $p=15$ (thick line).}
\end{figure}

In order to study the behavior of GCQW for large $t$, 
let us calculate the probabilities $P(k T)$ of multiple recurrences.  
Approximating $\zeta^k_{ju}$ (up to the second order term) by  
$\zeta^k_{ju} =\pm \exp(2i(-1)^uk\lambda_j)$ leads to  
\begin{equation}\label{eqkt}
P(kT)=J_0^2(2k\sqrt{\Deff})\,,
\end{equation} 
where $J_0(x)$  
is a Bessel function (Fig.~\ref{f2}).
Thus, the probability $P(kT)$ does not decrease monotonically with $\Deff$ 
but oscillates with this parameter, converging to $0$ as $k\to\infty$. 
Since the first zero of $J_0(x)$ occurs at $x\sim 2.4$, 
the decay of recurrences can be roughly characterized by time   
\begin{equation}\label{eqtau}
\tau(p)\sim 1.2\; T \Deff^{-1/2}\,.
\end{equation} 
This fact explains nicely numerical simulations for $D=1/2$ 
given by Buerschaper and Burnett~\cite{Bue} who, instead of 
recurrences, analysed the behavior of the related characteristic of 
a random walk: its standard deviation $\sigma$. 
For example $\tau(10)\sim 67.8$, while $\tau(20)\sim 768$.  
Hence, the oscillations of $\sigma$ for $p=10$ die quickly (Fig.~2b in~\cite{Bue}), 
while for $p=20$ they can still be seen for $t\sim200$  (see Fig.~2c in~\cite{Bue}). 
In Fig.~\ref{f3} we show the striking difference in $\sigma$ dynamics for odd and even 
values of $p$. 
In the even case, when $\Deff=D^{p/2}$,  (e.g., $\tau(16)\sim 307.2$), 
oscillations of $\sigma$ are damped more efficiently 
then in the odd case in which $\Deff=D^{p}$ ($\tau(15)\sim 6516.7$). 
Numerical simulations suggest that for  $t\gg\tau(p)$ a ballistic diffusion 
occurs with $\sigma$ proportional to $t$. In fact, the system behaves 
similarly to the basic model of CQW with $D$ replaced by $\Deff$; 
for instance, the formula $\sigma=t\sqrt{1-\sqrt{1-D}}$ of Konno~\cite{Kon}
in the case $\Phi=2\pi\frac{q}{p}$ becomes 
\begin{equation}\label{eqsigma}
\sigma=t\sqrt{1-\sqrt{1-\Deff}}\,.
\end{equation}  
This formula holds for both odd and even $p$, although, as we have just
observed,  because of difference in 
$\Deff$, the value of $\sigma/t$ for $p$ and $p+1$ can be dramatically different.

\begin{figure}
\scalebox{.8}
{\includegraphics{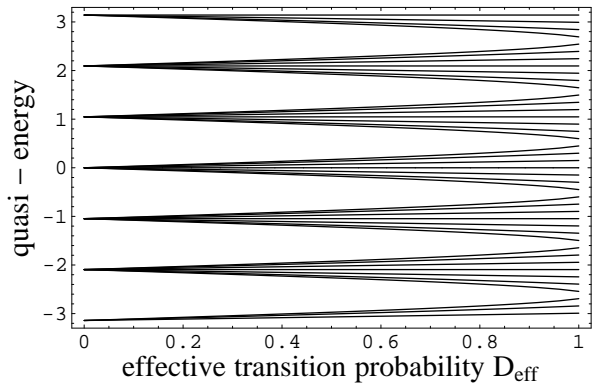}}
\caption{\label{f4} The levels of quasi-energy as a function of $\Deff$ for
$p=3$, $N=21$.}
\end{figure}

One way to explain the behavior of GCQW described above is to compare this system 
to one with a 1D periodic potential.  
The nodes of the $N$-cycle correspond to potential wells. 
Since we use two sided coin in the corresponding system 
we have two bands of Bloch levels, each consisting of $N$ levels, and
the width of each band proportional to the tunnelling amplitude~$d$. 
For $\Phi=0$ the symmetry of the system is given by $\mathbb Z_N$. 
If $\Phi=2\pi \tfrac{q}{p}$ 
the symmetry group reduces to $\mathbb Z_{N/p}$ and 
one can view such a system as a family of $N/p$ clusters of wells, 
each with $2p$ energy levels, 
and the tunnelling amplitude $\deff=d^p$, 
(so, the width of the new band is proportional to $\deff$).  
The quasi-energies $E_{jus}$ of the system are given by $\textrm{Arg}(r_{jus})$
and, since the system evolves in discrete time, 
the whole spectrum of quasi-energies is contained within range of $2\pi$. 
Thus, at $\deff=1$, the width of a single band is $2\pi/2p$. 
Consequently, for $\deff <1$, the width of each band should be approximately 
$\tfrac{2\pi}{2p}\deff$. 
Our analytical computations agree with this heuristic perfectly (Fig.~\ref{f4}).
This picture is valid only for $p$ odd; when $p$ is even an 
additional degeneration emerges and the resulting system has only $p$ 
quasi-energy bands; so, in this case, $\deff=d^{p/2}$. 
In both cases, if $\deff$ is small, the energy bands  are very narrow, 
the resulting system is nearly harmonic 
and a periodic behaviour is expected. 
On the other hand, due to the uncertainty relation, 
the non-harmonic effects are expected to emerge only  
after time $t$ larger than the reciprocity of the band width.
Note that it is consistent with $\tau(p)$ given by (\ref{eqtau}).

Let us remark that although we study the case $\Phi=2\pi \tfrac{q}{p}$ 
none of the results presented so far have depended on $q$.  
The reason for that is that we have considered only "almost perfect" recurrences, 
considering only times $t$ which are multiplicities of $T$. 
However, one can also observe ``imperfect'' recurrences for which 
the probability of recurrence $P(t)$ has a visible maximum, 
yet typically is smaller than for the  perfect ones. 
The existence of such an imperfect occurrences is caused 
by the fact that our system is discrete, 
whereas in the continuous version of the problem the maximum of $P(t)$ might 
occur at $t$ which is not an integer. 
From our previous considerations it comes as no surprise that an excellent 
candidate for a continuous  version of our dynamics is a Bloch system. 
To make this analogy more transparent, consider an equivalent, more symmetric, 
coin operator 
$C'=\sum_c\big(i d|c\rangle\langle c|+a |c\rangle\langle 1-c|\big)$,
and let $|\Psi(t)\rangle=\sum_{c,n} \alpha_{c,n}(t)|c\rangle|n\rangle$. 
Then the evolution equation $|\Psi(t)\rangle=U^t_\phi|\Psi(0)\rangle$  
leads to  the recursive  equation 
\begin{multline*}\label{eq5}
\alpha_{c,n}(t+1)-\alpha_{c,n}(t-1)=\Big(e^{i(2n-1)\Phi}-1\Big)\alpha_{c,n}(t-1)\\
+ide^{i(n-1)\Phi}(\alpha_{c,n-1}(t)+\alpha_{c,n+1}(t)\Big)\,,
\end{multline*}
for $c=0,1$. Note that in the case of $\Phi=0$ an analogous equation 
has been studied in \cite{Kni,Rom,Knig}. 
If $n \ll \tilde T \equiv 2\pi/\Phi$ this equation can be approximated by the differential 
equation
\begin{equation}\label{eq6}
\frac{d\alpha_{c,n}}{dt}=in\Phi\alpha_{c,n}+\frac{id}{2}(\alpha_{c,n-1}+\alpha_{c,n+1})\,,
\end{equation}
which can be identified with the coupled mode equations describing optical Bloch 
oscillations observed in a waveguide array with linearly growing effective index 
\cite{Per,Mor}.  
Thus, the recurrences in the GCQW dynamics can be understood as a discrete version of  
optical Bloch oscillations. 
The exact solution of (\ref{eq6}) (see~\cite{Per}) gives     
\begin{equation}\label{eq7}
P(t)=J^2_0\Big[\frac{d\,\tilde T}{\pi}\sin\Big(\frac{\pi t}{\tilde T}\Big)\Big]\,.
\end{equation}
Thus, a prefect recurrence should be observed for $t=k \tilde T$. 
In a discrete version, for $t=k\tilde T$ and 
$t\ll \tau(p)$, we indeed observe a (nearly) perfect recurrence 
provided $k\tilde T$ is an even integer (for each odd $t$ we have  $P(t)=0$); 
if this is not the case, 
then an imperfect recurrence occurs at one of the neighboring steps. 
A numerical simulation fully confirms these anticipations.
 
	Finally, let us comment on the case when $\tilde T=2 \pi/\Phi$ is irrational.
Then,  $\tilde T=\lim \tfrac{p_n}{q_n}$, with  $p_n$ (and $q_n$)  tending to infinity as 
$n \to \infty$. As we have shown above the ballistic diffusion can take place 
only after time $\tau(p_n)$, which in this case goes  to infinity, 
so the diffusion will be suppressed forever. 
Moreover, we can expect that the probability that the particle returns 
to the initial state in time $t$ oscillates irregularly with period $\tilde T$.
One can also easily see that the maximum value $\sigma_{\textrm{max}}$ 
of the standard deviation (localization length) can be approximated by 
\begin{equation}\label{eqmax}
\sigma_{\textrm{max}}=\frac{\tilde T}{2}\sqrt{1-\sqrt{1-D}}\,.
\end{equation}
Again, this agrees well with earlier simulations for  $D=1/2$ and $\tilde T=4\pi$   
given by Buerschaper and Burnett \cite{Bue} 
(see Fig.~2d therein, for which $\sigma_{\textrm{max}}=3.4$).
 
In the paper we have studied the behavior of GCQW, 
where the phase is modified by $\Phi=2\pi \tfrac{q}{p}$
at each mode. For such a system the probabilities of recurrences and multiple recurrences 
are given by (\ref{eq2}) and (\ref{eqkt}) respectively. 
They depend strongly on the parity of $p$ which can be attributed to
a ``degeneration'' of quasi-energy levels in the case when  $p$ is even. 
We showed that the system oscillates for 
small times $t$ while for $t$  much larger than $\tau(p)$, (see (\ref{eqtau})), 
a ballistic diffusion occurs for every $D<1$, with the standard deviation $\sigma$ 
given by (\ref{eqsigma}). For an irrational $\Phi/2\pi$, our model predicts
dynamical localization and gives a good upper bound  
for the localization length $\sigma_{\textrm{max}}$
(see~(\ref{eqmax})).
Our results are in perfect agreement with experimental and numerical observations from 
\cite{Bou,Bue} but some of their consequences are still to  be verified  experimentally 
(a natural candidate for an experimental realization of GCQW 
would be a version of spin-dependent transport of atoms in optical 
lattices~\cite{Man,Mand}).

The authors  wish to thank the State Committee for Scientific Research
(KBN) for its support: 
 A.W.\ and A.G.\ were supported 
by grant 0~T00A~003~23;
T.\L.\ and M.B. by grant 1 P03A 025 27. 


\begin{thebibliography}{99}
\expandafter\ifx\csname natexlab\endcsname\relax\def\natexlab#1{#1}\fi
\expandafter\ifx\csname bibnamefont\endcsname\relax
  \def\bibnamefont#1{#1}\fi
\expandafter\ifx\csname bibfnamefont\endcsname\relax
  \def\bibfnamefont#1{#1}\fi
\expandafter\ifx\csname citenamefont\endcsname\relax
  \def\citenamefont#1{#1}\fi
\expandafter\ifx\csname url\endcsname\relax
  \def\url#1{\texttt{#1}}\fi
\expandafter\ifx\csname urlprefix\endcsname\relax\def\urlprefix{URL }\fi
\providecommand{\bibinfo}[2]{#2}
\providecommand{\eprint}[2][]{\url{#2}}


\bibitem[{\citenamefont{Moo}(1995)}]{Moo}
\bibinfo{author}{\bibfnamefont{F.L.}~\bibnamefont{Moore {\it et al.}}},
  \bibinfo{journal}{Phys.\ Rev.\ Lett.} \textbf{\bibinfo{volume}{75}},
  \bibinfo{pages}{4598} (\bibinfo{year}{1995}).

\bibitem[{\citenamefont{Amm}(1998)}]{Amm}
\bibinfo{author}{\bibfnamefont{H.}~\bibnamefont{Ammann}},
\bibinfo{author}{\bibfnamefont{R.}~\bibnamefont{Gray}},
\bibinfo{author}{\bibfnamefont{I.}~\bibnamefont{Shvarchuck}},
\bibinfo{author}{\bibfnamefont{N.}~\bibnamefont{Christensen}},
  \bibinfo{journal}{Phys.\ Rev.\ Lett.} \textbf{\bibinfo{volume}{80}},
  \bibinfo{pages}{4111} (\bibinfo{year}{1998}).

\bibitem[{\citenamefont{Ober}(1999)}]{Ober}
\bibinfo{author}{\bibfnamefont{M.K.}~\bibnamefont{Oberthaler {\it et al.}}},
  \bibinfo{journal}{Phys.\ Rev.\ Lett.} \textbf{\bibinfo{volume}{83}},
  \bibinfo{pages}{4447} (\bibinfo{year}{1999}).

\bibitem[{\citenamefont{Arcy}(2001)}]{Arcy}
\bibinfo{author}{\bibfnamefont{M.B.}~\bibnamefont{d'Arcy {\it et al.}}},
  \bibinfo{journal}{Phys.\ Rev.\ Lett.} \textbf{\bibinfo{volume}{87}},
  \bibinfo{pages}{74102} (\bibinfo{year}{2001}).

\bibitem[{\citenamefont{Mon}(2004)}]{Mon}
\bibinfo{author}{\bibfnamefont{S.}~\bibnamefont{Montangero}},
\bibinfo{author}{\bibfnamefont{A.}~\bibnamefont{Romito}},
\bibinfo{author}{\bibfnamefont{G.}~\bibnamefont{Benenti}},
\bibinfo{author}{\bibfnamefont{R.}~\bibnamefont{Fazioo}},
\eprint{cond-mat/0407274}. 

\bibitem[{\citenamefont{Bou}(1999)}]{Bou}
\bibinfo{author}{\bibfnamefont{D.}~\bibnamefont{Bouwmeester {\it et al.}}},
  \bibinfo{journal}{Phys.\ Rev.\ A} \textbf{\bibinfo{volume}{61}},
  \bibinfo{pages}{13410} (\bibinfo{year}{1999}).

\bibitem[{\citenamefont{Har}(1997)}]{Har}
\bibinfo{author}{\bibfnamefont{D.A.}~\bibnamefont{Harmin}},
  \bibinfo{journal}{Phys.\ Rev.\ A} \textbf{\bibinfo{volume}{56}},
  \bibinfo{pages}{232} (\bibinfo{year}{1997}).

\bibitem[{\citenamefont{Tor}(1998)}]{Tor}
\bibinfo{author}{\bibfnamefont{P.}~\bibnamefont{T\"orm\"a}},
  \bibinfo{journal}{Phys.\ Rev.\ Lett.} \textbf{\bibinfo{volume}{81}},
  \bibinfo{pages}{2185} (\bibinfo{year}{1998}).

\bibitem[{\citenamefont{Kni}(2003)}]{Kni}
\bibinfo{author}{\bibfnamefont{P.L.}~\bibnamefont{Knight}},
\bibinfo{author}{\bibfnamefont{E.}~\bibnamefont{Rold\'an}},
\bibinfo{author}{\bibfnamefont{J.E.}~\bibnamefont{Sipe}},
  \bibinfo{journal}{Phys.\ Rev.\ A} \textbf{\bibinfo{volume}{68}},
  \bibinfo{pages}{20301(R)} (\bibinfo{year}{2003}).

\bibitem[{\citenamefont{ADZ}(19933)}]{ADZ}
\bibinfo{author}{\bibfnamefont{Y.}~\bibnamefont{Aharonov}},
\bibinfo{author}{\bibfnamefont{L.}~\bibnamefont{Davidovich}},
\bibinfo{author}{\bibfnamefont{N.}~\bibnamefont{Zagury}},
  \bibinfo{journal}{Phys.\ Rev.\ A} \textbf{\bibinfo{volume}{48}},
  \bibinfo{pages}{1687} (\bibinfo{year}{1993}).

\bibitem[{\citenamefont{Kem}({\natexlab{a}})}]{Kem}
\bibinfo{author}{\bibfnamefont{J.}~\bibnamefont{Kempe}},
\bibinfo{journal}{Contemporary Phys.} \textbf{\bibinfo{volume}{44}},
  \bibinfo{pages}{307} (\bibinfo{year}{2003}).

\bibitem[{\citenamefont{Bue}(2003)}]{Bue}
\bibinfo{author}{\bibfnamefont{O.}~\bibnamefont{Buerschaper}},
\bibinfo{author}{\bibfnamefont{K.}~\bibnamefont{Burnett}},
  \eprint{quant-ph/0406039}.

\bibitem[{\citenamefont{Aha}(2001)}]{Aha}
\bibinfo{author}{\bibfnamefont{D.}~\bibnamefont{Aharonov}},
  \bibinfo{author}{\bibfnamefont{A.}~\bibnamefont{Ambainis}},
  \bibinfo{author}{\bibfnamefont{J.}~\bibnamefont{Kempe}}, \bibnamefont{and}
  \bibinfo{author}{\bibfnamefont{U.}~\bibnamefont{Vazirani}},
  \bibinfo{journal}{Proc. of the 30th Annual ACM Symposium on Theory of
  Computation (ACM Press, New York, 2001)}
 \bibinfo{pages}{50} (\bibinfo{year}{2001}).

\bibitem[{\citenamefont{Kon}(2003)}]{Kon}
\bibinfo{author}{\bibfnamefont{N.}~\bibnamefont{Konno}},
\bibinfo{journal}{Quantum Information Processing} \textbf{\bibinfo{volume}{1}},
  \bibinfo{pages}{345} (\bibinfo{year}{2003}).

\bibitem[{\citenamefont{Rom}(2004)}]{Rom}
\bibinfo{author}{\bibfnamefont{A.}~\bibnamefont{Romanelli {\em et al.}}},
  \bibinfo{journal}{Physica A} \textbf{\bibinfo{volume}{338}},
  \bibinfo{pages}{395} (\bibinfo{year}{2004}).

\bibitem[{\citenamefont{Knig}(2004)}]{Knig}
\bibinfo{author}{\bibfnamefont{P.L.}~\bibnamefont{Knight}},
\bibinfo{author}{\bibfnamefont{E.}~\bibnamefont{Rold\'an}},
\bibinfo{author}{\bibfnamefont{J.}~\bibnamefont{Sipe}},
  \eprint{quant-ph/0312133}.

\bibitem[{\citenamefont{Per}(1999)}]{Per}
\bibinfo{author}{\bibfnamefont{T.}~\bibnamefont{Pertsch {\em et al.}}},
  \bibinfo{journal}{Phys.\ Rev.\ Lett.} \textbf{\bibinfo{volume}{83}},
  \bibinfo{pages}{4752} (\bibinfo{year}{1999}).

\bibitem[{\citenamefont{Mor}(1999)}]{Mor}
\bibinfo{author}{\bibfnamefont{R.}~\bibnamefont{Morandotti {\em et al.}}},
  \bibinfo{journal}{Phys.\ Rev.\ Lett.} \textbf{\bibinfo{volume}{83}},
  \bibinfo{pages}{4756} (\bibinfo{year}{1999}).

\bibitem[{\citenamefont{Man}(2003)}]{Man}
\bibinfo{author}{\bibfnamefont{O.}~\bibnamefont{Mandel {\em et al.}}},
  \bibinfo{journal}{Phys.\ Rev.\ Lett.} \textbf{\bibinfo{volume}{91}},
  \bibinfo{pages}{010407} (\bibinfo{year}{2003}).

\bibitem[{\citenamefont{Mand}(2003)}]{Mand}
\bibinfo{author}{\bibfnamefont{O.}~\bibnamefont{Mandel {\em et al.}}},
  \bibinfo{journal}{Nature} \textbf{\bibinfo{volume}{425}},
  \bibinfo{pages}{937} (\bibinfo{year}{2003}).

\end{thebibliography}
\end{document}